\documentclass[letter,longauth]{aa} 

\usepackage{graphicx}
\usepackage{txfonts}
\usepackage[colorlinks=true, allcolors=blue]{hyperref}
\usepackage{orcidlink}
\usepackage{adjustbox}
\usepackage{multicol}
\usepackage{multirow}
\usepackage{xspace}
\newcommand {\ixpe}{\text{IXPE}\xspace}

\newcommand{\lumcgs}{erg\,s$^{-1}$}

\begin{document}

\title{Ultrasoft state of microquasar Cygnus X-3: \\ X-ray polarimetry reveals the geometry of astronomical puzzle}

\titlerunning{Cyg X-3 in ultrasoft state: a clue to the astronomical puzzle}

\author{Alexandra~Veledina\inst{\ref{in:UTU},\ref{in:Nordita}}\thanks{E-mail: alexandra.veledina@gmail.com}\orcidlink{0000-0002-5767-7253}
\and Juri~Poutanen \inst{\ref{in:UTU}}\orcidlink{0000-0002-0983-0049}
\and Anastasiia~Bocharova \inst{\ref{in:UTU}}
\and Alessandro~Di~Marco \inst{\ref{in:INAF-IAPS}}\orcidlink{0000-0003-0331-3259}
\and Sofia~V.~Forsblom \inst{\ref{in:UTU}}\orcidlink{0000-0001-9167-2790}
\and Fabio~La~Monaca \inst{\ref{in:INAF-IAPS},\ref{in:UniRoma2},\ref{in:LaSapienza}}\orcidlink{0000-0001-8916-4156} 
\and Jakub~Podgorn\'{y} \inst{\ref{in:CAS-ASU}}\orcidlink{0000-0001-5418-291X}
\and Sergey~S.~Tsygankov \inst{\ref{in:UTU}}\orcidlink{0000-0002-9679-0793} 
\and Andrzej~A.~Zdziarski \inst{\ref{in:CAMK}}\orcidlink{0000-0002-0333-2452} 
\and Varpu Ahlberg\inst{\ref{in:UTU}}\orcidlink{0009-0006-9714-5063}
\and David~A.~Green \inst{\ref{in:UnivCam}}\orcidlink{0000-0003-3189-9998} 
\and Fabio~Muleri \inst{\ref{in:INAF-IAPS}}\orcidlink{0000-0003-3331-3794} 
\and Lauren~Rhodes\inst{\ref{in:UnivOxf}}\orcidlink{0000-0003-2705-4941} 
\and Stefano~Bianchi \inst{\ref{in:Roma3}}\orcidlink{0000-0002-4622-4240}
\and Enrico~Costa \inst{\ref{in:INAF-IAPS}}\orcidlink{0000-0003-4925-8523}
\and Michal~Dov\v{c}iak \inst{\ref{in:CAS-ASU}}\orcidlink{0000-0003-0079-1239}
\and Vladislav Loktev\inst{\ref{in:UTU}}\orcidlink{0000-0001-6894-871X}
\and Michael~McCollough \inst{\ref{in:CfA}}\orcidlink{0000-0002-8384-3374} 
\and Paolo~Soffitta \inst{\ref{in:INAF-IAPS}}\orcidlink{0000-0002-7781-4104}
\and Rashid~Sunyaev \inst{\ref{in:MPA},\ref{in:IKI}}\orcidlink{0000-0002-2764-7192}
}

\authorrunning{A. Veledina et al.}

\institute{
Department of Physics and Astronomy, FI-20014 University of Turku, Finland \label{in:UTU}
\and
Nordita, KTH Royal Institute of Technology and Stockholm University, Hannes Alfv\'ens v\"ag 12, SE-10691 Stockholm, Sweden \label{in:Nordita}
\and INAF Istituto di Astrofisica e Planetologia Spaziali, Via del Fosso del Cavaliere 100, 00133 Roma, Italy \label{in:INAF-IAPS}
\and Dipartimento di Fisica, Universit\`{a} degli Studi di Roma ``Tor Vergata'', Via della Ricerca Scientifica 1, 00133 Roma, Italy \label{in:UniRoma2}
\and Dipartimento di Fisica, Universit\`{a} degli Studi di Roma ``La Sapienza'', Piazzale Aldo Moro 5, 00185 Roma, Italy  \label{in:LaSapienza}
\and Astronomical Institute of the Czech Academy of Sciences, Boční II 1401/1, 14100 Praha 4, Czech Republic \label{in:CAS-ASU} 
\and Nicolaus Copernicus Astronomical Center, Polish Academy of Sciences, Bartycka 18, PL-00-716 Warszawa, Poland \label{in:CAMK}
\and Astrophysics Group, Cavendish Laboratory, 19 J. J. Thomson Avenue, Cambridge, CB3 0HE, UK \label{in:UnivCam} 
\and Harvard-Smithsonian Center for Astrophysics, 60 Garden St, Cambridge, MA 02138, USA \label{in:CfA}
\and Astrophysics, Department of Physics, University of Oxford, Denys Wilkinson Building, Keble Road, Oxford, OX1 3RH, UK \label{in:UnivOxf}
\and Dipartimento di Matematica e Fisica, Universit\`{a} degli Studi Roma Tre, Via della Vasca Navale 84, 00146 Roma, Italy \label{in:Roma3}
\and Max Planck Institute for Astrophysics, Karl-Schwarzschild-Str 1, D-85741 Garching, Germany \label{in:MPA}
\and Space Research Institute, Russian Academy of Sciences, Profsoyuznaya 84/32, 117997 Moscow, Russia \label{in:IKI}
}

\date{Received ...; Accepted ...}

 
\abstract
{Cygnus X-3 is an enigmatic X-ray binary, that is both an exceptional accreting system and a cornerstone for the population synthesis studies.
Prominent X-ray and radio properties follow a well-defined pattern, yet the physical reasons for the state changes observed in this system are not known.
Recently, the presence of an optically thick envelope around the central source in the hard state was revealed using the X-ray polarization data obtained with Imaging X-ray Polarimetry Explorer (\ixpe). 
In this work, we analyse \ixpe data obtained in the ultrasoft (radio quenched) state of the source. 
The average polarization degree (PD) of $11.9\pm0.5\%$ at a polarization angle (PA) of $94\degr\pm1\degr$ is inconsistent with the simple geometry of the accretion disc viewed at an intermediate inclination.
The high PD, the blackbody-like spectrum, 
and the weakness of fluorescent iron line imply that the central source is hidden behind the optically thick outflow and its beamed radiation is scattered towards our line of sight.
In this picture the observed PD is directly related to the source inclination, which we conservatively determine to lie in the range $26\degr<i<28\degr$. 
.pdfUsing the new polarimetric properties, we propose the scenario that can be responsible for the cyclic behaviour of the state changes in the binary.}


   \maketitle
%

\section{Introduction}
\label{sec:intro}

Accreting binary system Cygnus X-3 (hereafter Cyg X-3) is one of the first detected X-ray sources \citep{Giacconi1967}.
This persistent source shows several exceptional observational properties. 
It is the brightest X-ray binary in radio wavelengths, with fluxes that can reach 20~Jy \citep{McCollough1999,Corbel2012}, and is one of a few Galactic binaries with detectable $\gamma$-ray emission \citep{Tavani2009,Atwood2009}.
It is one of the shortest-period X-ray binaries, $P_{\rm orb}=4.8$~h, that is also rapidly changing owing to the high mass loss rate from the system  \citep{vanderKlis1981,Antokhin2019}.
Short period implies that the luminous companion star is very compact, suggesting that the donor is a Wolf-Rayet (WR) star, which is further supported by the hydrogen-depleted infrared spectrum  \citep{vanKerkwijk1992,vanKerkwijk1996,Fender1999}.
This unique binary plays an important role in population synthesis studies: a successful model is obliged to reproduce the presence of Cyg~X-3 in the Milky Way, the compact object in a tight binary orbit with the WR companion; at the same time, current estimates suggest the presence of only one such system in our Galaxy \citep{Lommen2005}.

\begin{figure*}
\centering
\includegraphics[width=0.438\linewidth]{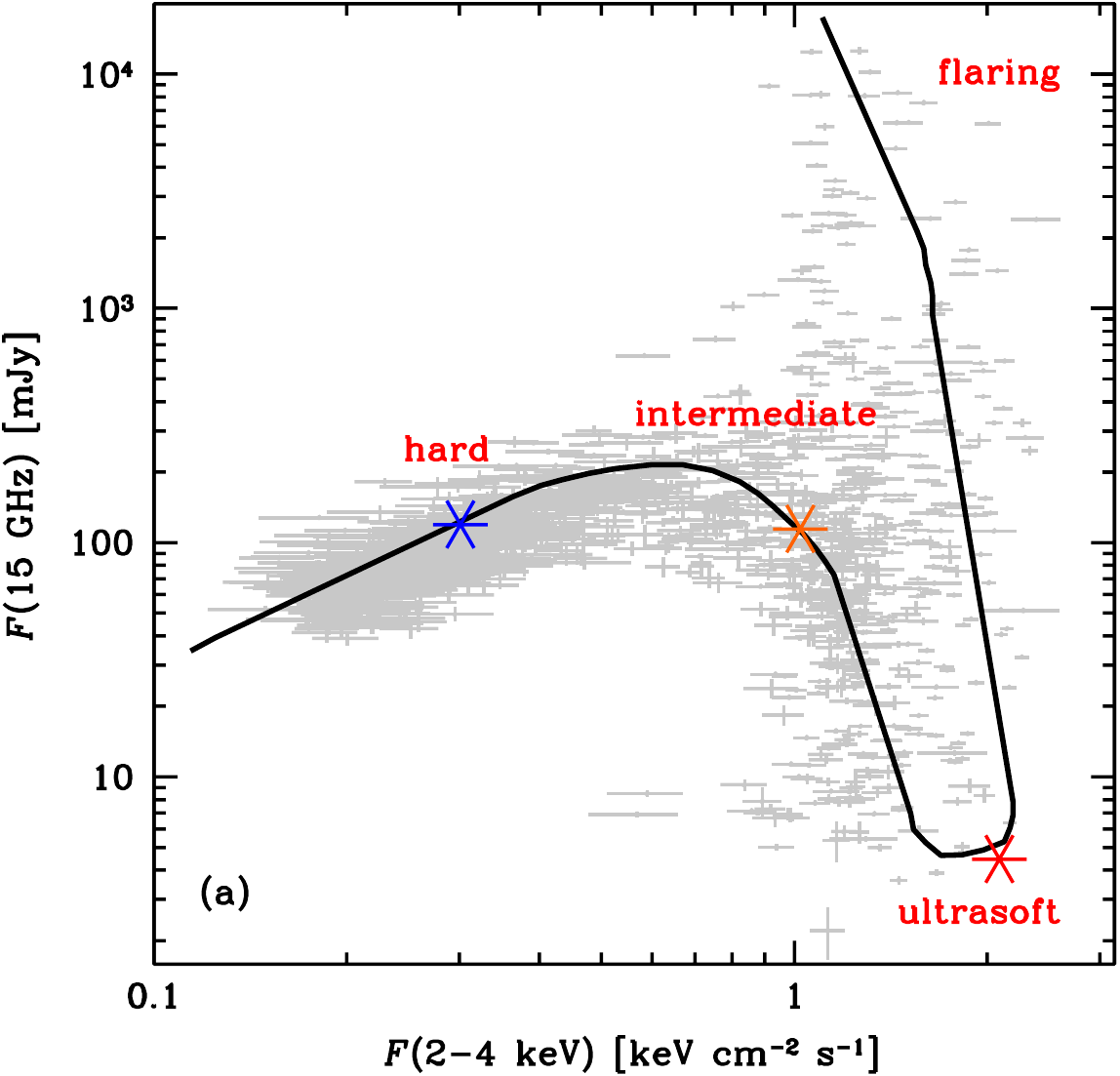}
\hspace{1cm}
\includegraphics[width=0.45\linewidth]{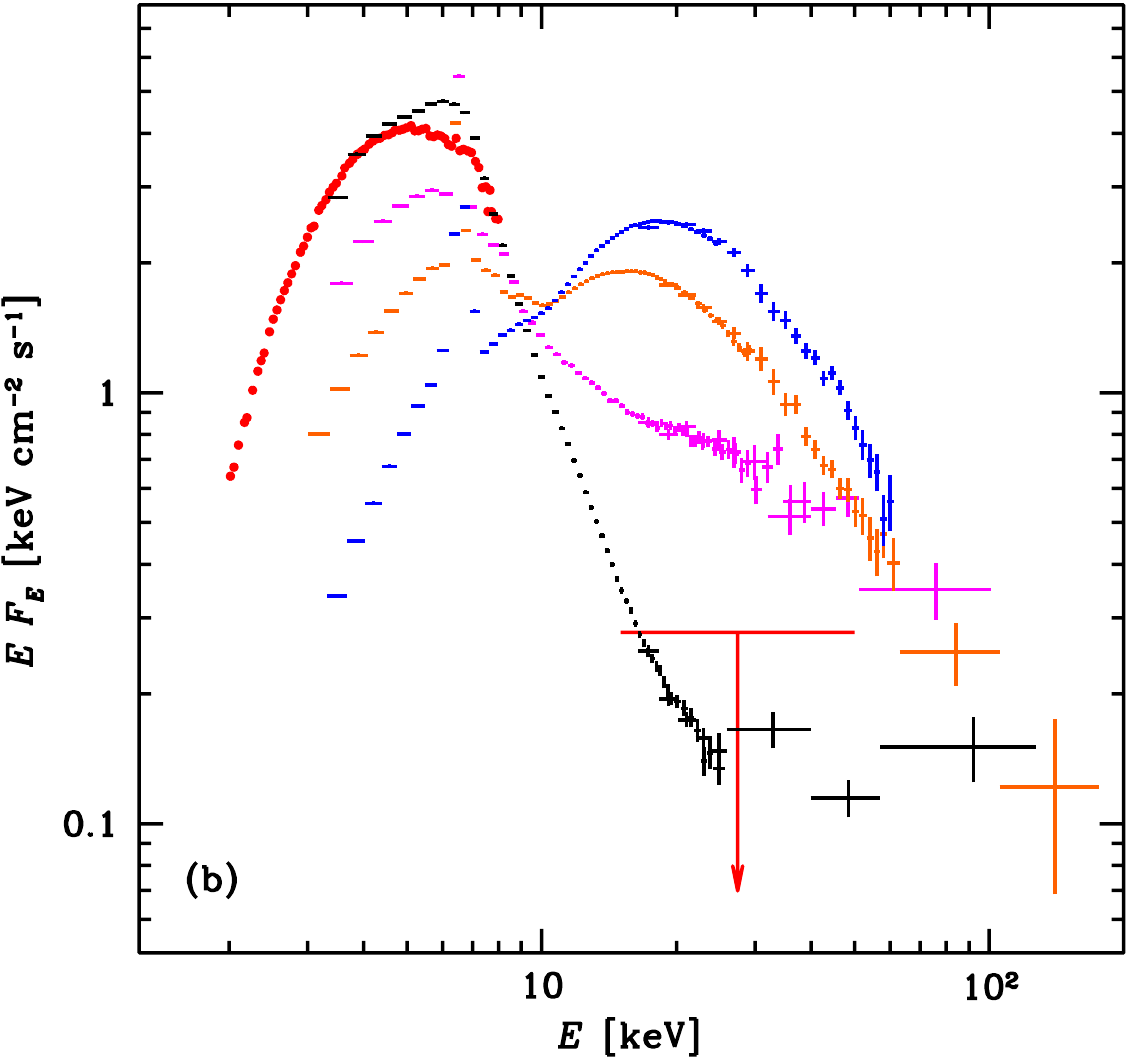}
\caption{Panel (a): Radio-X-ray track made by Cyg~X-3 as it swings between different spectral states (data from \citealt{Zdziarski2016}).
Panel (b): a set of average broadband spectra corresponding to the hard/radio quiescent (blue), intermediate/minor flaring (orange), ultrasoft/radio quenched (black), and soft non-thermal/flaring  (magenta) states as introduced in \citet{Szostek2008_states}. \ixpe data are shown with red symbols and the 2$\sigma$ upper limit at 30~keV (in red) is from Swift/BAT.
}
\label{fig:diagram_spectra}
\end{figure*}

\begin{figure}
\centering
\includegraphics[width=0.85\linewidth]{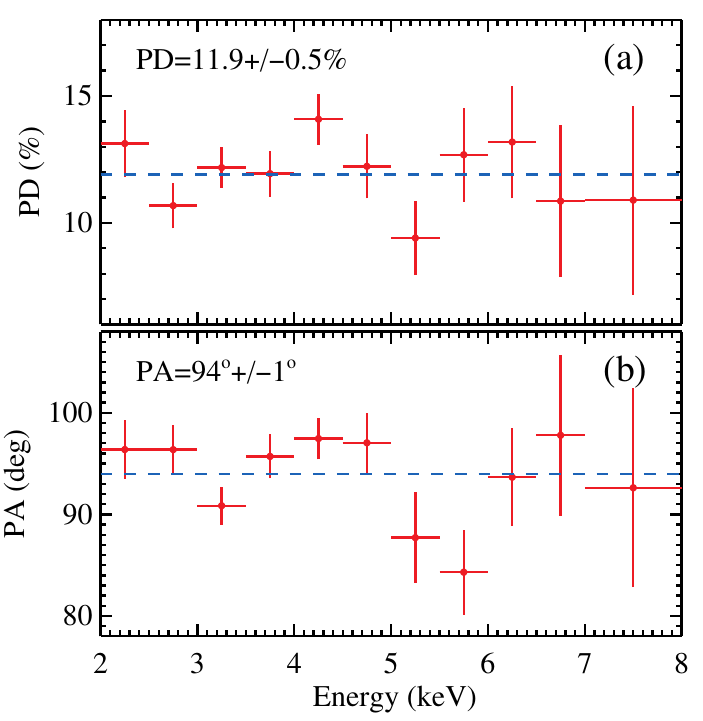} 
\caption{Average PD and PA as a function of energy. 
The horizontal dashed lines correspond the energy-average values. 
}
\label{fig:average_pdpa} 
\end{figure} 

There is no detected optical counterpart due to high distance to the source \citep[$D = 9.7\pm0.5$~kpc;][]{ReidMiller-Jones2023} and its location close to the galactic plane resulting in high absorption along to the line of sight.
However, the orbital phase dependence of radio, infrared, X-ray and $\gamma$-ray properties of the system have been studied extensively (e.g., \citealt{McCollough1999,Fender1999,Zdziarski2018}).
The source was found to regularly switch between several spectral states, characterised by distinct X-ray and radio properties \citep[Fig.~\ref{fig:diagram_spectra} and][]{Szostek2008_states,Koljonen2010}.
Transitions between the spectral states are generally thought to be related to the changes of accretion geometry in the system, but the exact picture of plasma configuration and understanding of the physical reasons for these changes are still missing. 
This makes the source an ideal target for X-ray polarimetric studies.

The source is most often found in the hard X-ray, quiescent radio state \citep{Waltman1996,Hjalmarsdotter2009}, when the X-ray continuum can be roughly described as a cutoff power law with a peak at around 20~keV (Fig.~\ref{fig:diagram_spectra}b), and prominent iron lines can be clearly identified.
Spectral decomposition in this state is complicated by the unknown absorption, in particular its component that is intrinsic to the source \citep{Hjalmarsdotter2008,Zdziarski2010}.

Cyg X-3 is found to make regular transitions towards the (ultra-)soft state, when the hard X-ray flux drops dramatically and the spectrum resembles a blackbody with $kT_{\rm bb}\approx 1.5$~keV.
Radio fluxes are likewise suppressed, motivating the name of the radio quenched state.
This transition is often followed by a major radio flare, when the highest radio fluxes are detected, while the X-ray continuum is moderately soft and exhibits high-energy tail \citep{Szostek2008_states,Koljonen2010}.

Exceptional multiwavelength properties of the source and fast, repeating swing between different states prompted \citet{Hjellming1973} to call it ``\textit{An astronomical puzzle}''.
Interest in solving this puzzle has persisted for more than 50 years. 
A new chapter in the study of the source has been opened with the launch of the Imaging X-ray Polarimetry Explorer (\ixpe), that enabled the first X-ray polarization studies in the 2--8~keV band.

The source was previously observed by \ixpe in the X-ray hard (radio quiescent) and intermediate (minor flaring) states \citep{Veledina2024CygX3}. 
The first X-ray polarimetric observations in October-November 2022 revealed high average polarization degree (PD) of $20.6\pm0.3$\% (becoming $23$\% after accounting for the unpolarized contribution of the iron line). 
The polarization angle, PA=$90\fdg1\pm0\fdg4$, suggests that it is nearly orthogonal to the extended jet emission and relativistic ejections observed in the system. 
Spectropolarimetric properties suggest the X-ray emission is produced solely by reflection, indicating the presence of obscuring material hiding the central engine from the observer's line of sight. 
The apparent high luminosity of $10^{38}$\,\lumcgs\ and a much higher luminosity emitted along the funnel in the obscuring outflow ($>5.5\times10^{39}$\,\lumcgs) imply that Cyg~X-3 belongs to the class of ultraluminous X-ray  sources.
The dramatic drop of polarization in the minor flaring state, PD=$10.0\pm0.5$\%, was interpreted in terms of the dilution of the outflow.

In this picture, the outflow could disappear completely in the ultrasoft and soft non-thermal states (radio quenched and major flaring states, respectively).
Here we present the X-ray polarimetric properties of \mbox{Cyg~X-3} in the ultrasoft state, with the aim to verify the source accretion geometry and to understand the physical reasons of state transitions.

\section{Data}

\ixpe observed the source in the ultrasoft state on 2024 June 2--3.
We follow the standard data analysis procedures that are described in detail in Appendix~\ref{app:ixpe}.
We use simultaneous Swift/BAT data\footnote{\url{https://swift.gsfc.nasa.gov/results/transients/}} \citep{Gehrels04} to estimate the flux in the hard X-ray band. 
We also monitored the source in radio  with the AMI telescope at 15~GHz. 
Results of radio monitoring Cyg X-3 are presented in Appendix~\ref{app:radio}.

\begin{figure}
\centering
\includegraphics[width=0.85\linewidth]{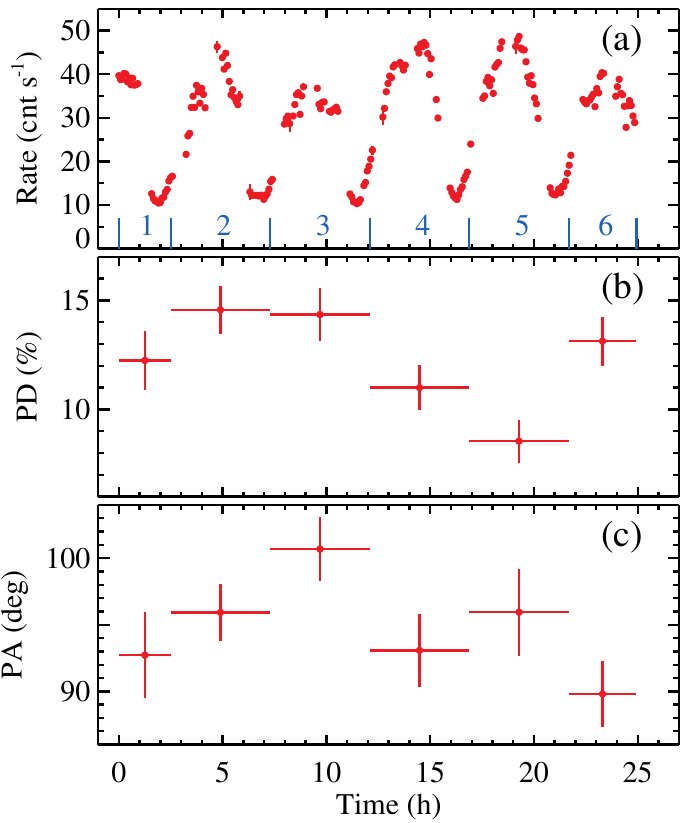} 
\caption{Time variations of the \ixpe count rate in the 2--8~keV band and the orbit-average PD and PA. 
Time is measured from the beginning of the observation at MJD~60463.8124. 
Boundaries of the time bins (except beginning and the end of the observation) correspond to the orbital phase 0.}
\label{fig:time_variab} 
\end{figure}

\section{Results}

The source was observed in the ultrasoft X-ray spectral state (Fig.~\ref{fig:diagram_spectra}b), with highly suppressed hard X-ray flux, as indicated by Swift/BAT.
We note the lower flux, with respect to its typical value \citep{Szostek2008_states}, around the iron line complex (compare the black and red spectra).
Radio fluxes were likewise historically low, $4.5\pm0.3$ mJy (Fig.~\ref{fig:diagram_spectra}a and Appendix~\ref{app:radio}).
The average polarimetric properties are shown in Fig.~\ref{fig:average_pdpa}. 
We find average PD=$11.9\pm0.5$\% and PA=$94\degr\pm1\degr$ and no significant energy dependence, in particular, around the iron line range.
The PA is shifted by $\sim$$4\degr$ with respect to the values detected in the hard and intermediate states.

Spectropolarimetric data suggest absence of the prominent iron line, in contrast to hard and intermediate states (see Appendix~\ref{app:ixpe}).
The spectrum in this state was previously well fitted either by a single blackbody component, a multicolour disc, or a low-temperature Comptonization component \citep{Szostek2008_states,Hjalmarsdotter2009,Koljonen2018}.
High PD and its independence of energy is, however, not consistent with either of the aforementioned components.

Next, we consider variations of polarimetric properties.
Prominent orbital flux variations are detected, with the typical profile shape of a brief dip and a more extended peak (Fig.~\ref{fig:time_variab}a).
In Fig.~\ref{fig:time_variab}bc we show the PD and PA averaged over different orbital cycles.
In contrast to the hard state, we find pronounced changes of PD in the ultrasoft state.
The PA remains constant within errors, yet systematically shifted with respect to the hard- and intermediate-state data in all bins.
We compared the spectropolarimetric properties of the time bins 2 and 3 (in Fig.~\ref{fig:time_variab}bc) to those of time bin 5 (see Figs.~\ref{fig:specpol_tbin23} and \ref{fig:specpol_tbin5} in Appendix~\ref{app:ixpe}).
The iron line is detected in the time bin 5, albeit at a low significance; for the bins 2+3 its contribution is compatible with zero.

\begin{figure}
\centering
\includegraphics[width=0.85\linewidth]{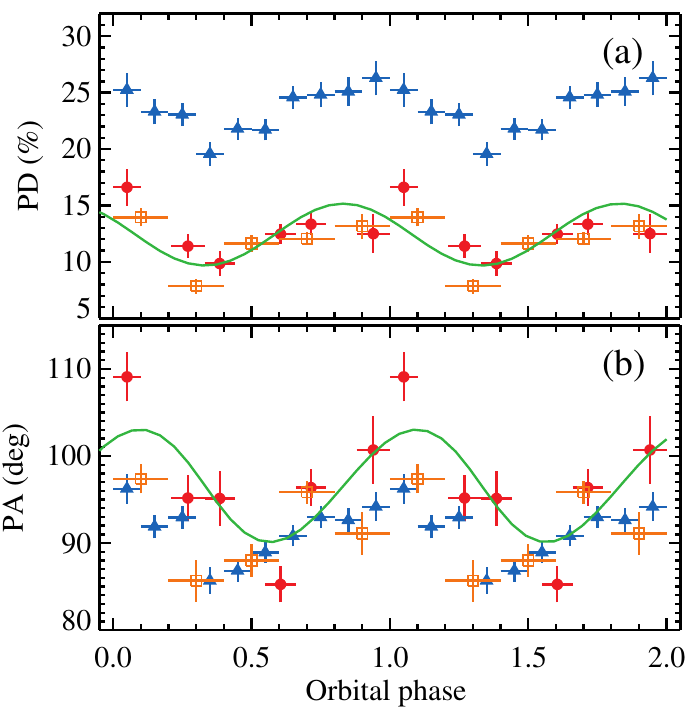}
\caption{Orbital phase dependence of the PD and PA from the hard state (blue triangles), the intermediate state (orange open squares), and ultrasoft state (red circles). 
Green lines show the best-fit rotating vector model (Appendix~\ref{app:rvm}) to the ultrasoft state data. 
}
\label{fig:PD_PA_orb} 
\end{figure}

Previous \ixpe observations in the hard and intermediate states revealed significant variations of polarization at the orbital period. 
In Fig.~\ref{fig:PD_PA_orb} we show the orbital phase-resolved polarimetric properties of the source in the three observed states.
The PD variations in the ultrasoft state are nearly identical to those found in the intermediate state, but the PA shows a pronounced peak around phase 0 (corresponding to the lowest X-ray fluxes).
Using the \texttt{pcube} analysis of the 2--5~keV and 5--8~keV energy bins, we also checked whether the energy dependence of PD is different for these two orbital phases.
We find no difference between the two energy bins for the high X-ray flux spectra, associated with phase 0.5, but for the low-flux bin (phase 0), the PD decreases with energy at 90\% confidence level.
We also performed orbital phase-resolved spectropolarimetry (see Figs.~\ref{fig:specpol_high} and \ref{fig:specpol_low}).
The iron line is detected in the spectra around phase 0; for the phase 0.5 only an upper limit can be put (see Appendix~\ref{app:ixpe}).

\begin{figure*}
\centering
\includegraphics[width=0.5\linewidth]{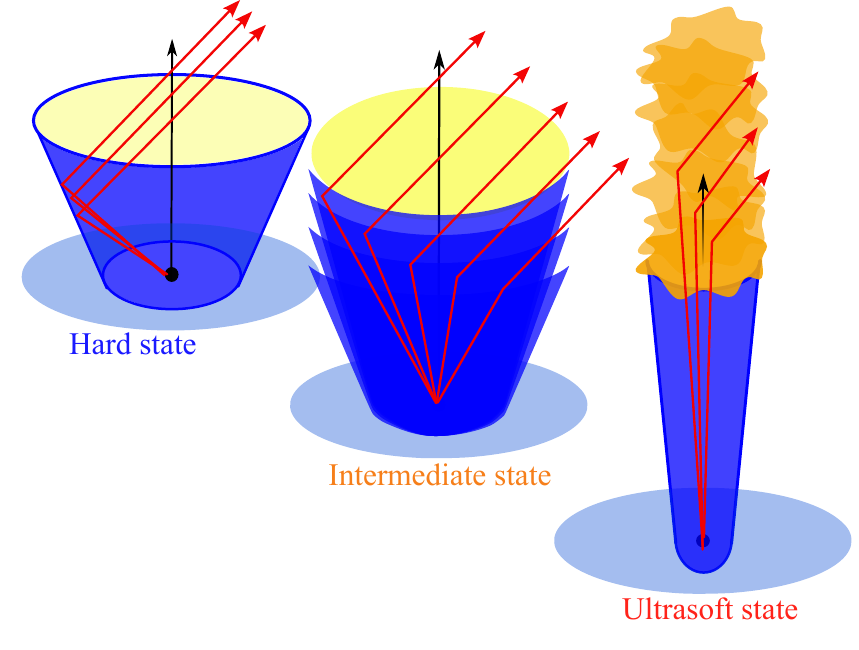} 
\includegraphics[width=0.4\linewidth]{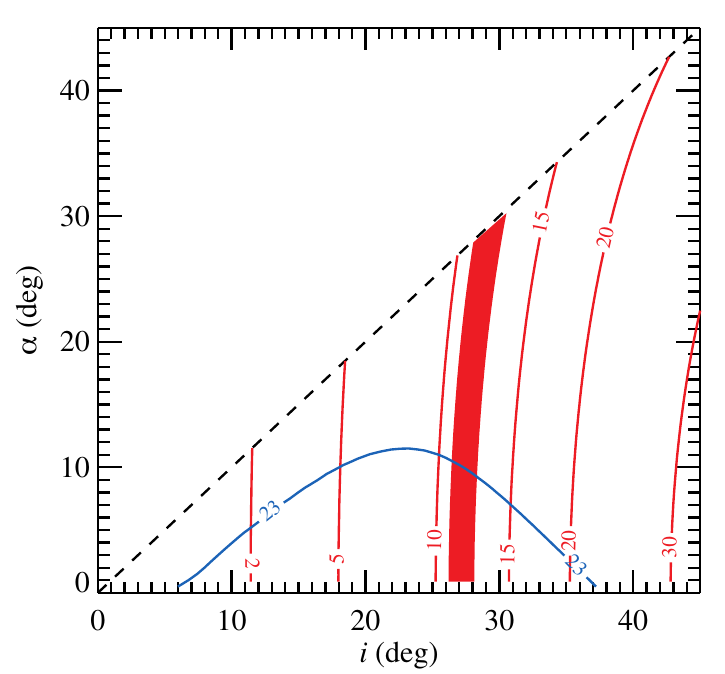}
\caption{Left: sketch of the geometry associated with the hard, intermediate, and ultrasoft states. 
Right: Polarization of radiation undergoing single scattering above a funnel of half-opening angle $\alpha$ for different observer inclinations $i$ (red lines). 
The filled red band depicts the conservative PD range 10.8--12.4\% of \mbox{Cyg~X-3} in the ultrasoft state which translates to the inclination of $i=$26\degr--28\degr, which is very close to $i=29\fdg5\pm1\fdg2$ determined by \citet{Antokhin2022}.
The blue line depicts the contour of the constant PD=23\% observed in the hard state for the model of the reflection from the inner wall of the conical surface \citep[from Fig.~4b in][]{Veledina2024CygX3}.}
\label{fig:pol_funnel} 
\end{figure*}

\section{Discussion}

Previous \ixpe studies clearly showed that the states of \mbox{Cyg X-3} do not simply match the hard/intermediate states of other X-ray binaries, and the standard scenarios that have been confirmed for, e.g., Swift~J1727.8$-$1613  \citep{Veledina2023,Ingram2024,Podgorny2024} and \mbox{Cyg~X-1} \citep{Krawczynski2022} cannot be applied to this source.
If the soft state of \mbox{Cyg X-3} is analogous to the other soft-state sources, the PD$\sim$1--2\% can be expected \citep[comparable to the measurements of \mbox{Cyg~X-1} and Swift~J1727.8$-$1613 that are thought to have similar orbital inclinations;][]{Svoboda2024,Steiner2024}.
On the contrary, we observe the average PD$\approx$12\%, that appears somewhat higher than that observed in the intermediate state.
This puts tight constraints on radiative mechanisms relevant to this state of the system, as well as the accretion geometry.

A high PD that is independent of energy is in line with the single reflection or scattering scenario.
Presence of an obscuring medium covering the central source along the line of sight is inevitable for the production of the high polarization.
Absence of a clear contribution of the iron line, both in spectra and polarimetric properties, indicates the spectrum is not dominated by the reflection off the cold and dense matter, suggested for the hard state \citep{Veledina2024CygX3}.
This conclusion is supported by the previous spectroscopic analysis of Chandra data \citep{Kallman2019}, that find absence of the fluorescent line from neutral iron (yet lines of highly ionised species of \ion{Fe}{xxv} and \ion{Fe}{xxvi}, whose motion is associated with the compact object, \citealt{Vilhu2009}, remain prominent throughout the ultrasoft state).

We have tested various scenarios where the observed polarization arises from the reflection at the inner walls of the funnel \citep[we use the same setup as described in][]{Podgorny2022,Podgorny2023,Veledina2024CygX3}.
We considered the orbital inclination in the range $20\degr<i<40\degr$, in line with previous estimates \citep{Vilhu2009,Antokhin2022}, and varied density and ionisation of the medium. 
When decreasing the density of the funnel walls (with respect to the hard-state $N_{\rm H}=10^{24}$~cm$^{-2}$), the PD becomes a decreasing function of energy in the \ixpe range, because the matter becomes effectively transparent to high-energy radiation.
We also considered cases of higher ionisation.
With increasing ionisation, we generally observe the decrease of the overall PD owing to the enhanced role of multiple scatterings.
We find that in order to reproduce the lack of the neutral/moderately ionized iron lines, the ionisation parameter (giving the ratio of ionising flux to the number density of matter) $\xi\gtrsim5000$ is required. 
This value is somewhat high, yet still within the boundaries of the allowed range $2\lesssim \log\xi \lesssim 4$ coming from the line spectroscopy \citep{Kallman2019}.
However, at these high values the multiple scatterings occur within a large fraction of the envelope, akin to a spherical shell, hence the PD drops down to 1\%, becoming inconsistent with the observed levels.

The observed blackbody-like spectrum and PD=10--15\% naturally appear in the scenario where the intrinsic radiation undergoes single Thomson scattering in the medium along the axis of the optically thick funnel and the observer sees only this scattered component (see Fig.~\ref{fig:pol_funnel}).
In this case, the observed average PD gives the direct estimate on the inclination.
For single Thomson scattering regime we have
\begin{equation}
\label{eq:pdmu} 
  P = \frac{1-\mu^2}{1+\mu^2},
\end{equation}
where $\mu$ is the cosine of the scattering angle. 
Thus PD=$11.9\pm0.5$\% translates to inclination $i=\arccos\left[(1-P)/(1+P)\right]^{1/2}=27\fdg5\pm 0\fdg5$. 
This constraint is relevant to a geometry of an infinitely narrow funnel. 
Calculations that include averaging over the non-zero opening angle $\alpha$ of a conical funnel, for an isotropic central source and equal scattering probability within the funnel, show that PD dependence on $\alpha$ is weak (see nearly vertical contours of constant PD in Fig.~\ref{fig:pol_funnel}).

Orbital variability can be used as an independent tracer of accretion geometry.
Orbital profiles of gamma-ray fluxes were previously explained by a jet-orbital axis misalignment \citep{Dubus2010,Zdziarski2018} or bending of the jet/funnel \citep{Dmytriiev2024}.
Radio fluxes were likewise varying due to changing wind absorption \citep{Zdziarski2018}.
Interestingly though, we find absence of the orbital modulation of radio fluxes during \ixpe observations (Fig.~\ref{fig:radio_ami2} in Appendix~\ref{app:radio}).
Polarization properties of the precessing funnel can be tested using the rotating vector model \citep[see Appendix~\ref{app:rvm};][]{Radhakrishnan1969,Poutanen2020}. 
We note that this model does not necessarily assume that the radiation beam collimated by the funnel is precessing, but there could be an asymmetry in the distribution of scattering material, e.g., formed by the bow shock \citep{Antokhin2022} with the geometry fixed in the corotating frame of the binary.
In Fig.~\ref{fig:PD_PA_orb}, we show the best-fit model with green line.
The fit gives $\chi^2$/d.o.f=25.7/8. 
The model fails in reproducing the data because the amplitude of PA variations in excess of $20\degr$ translates to the changes of PD, predicted by this model for $i=27\degr$, by about 9\%, much larger than the observed PD amplitude.
While the model can reproduce orbital variations of PA, precessing funnel scenario cannot be responsible for the orbital changes of PD. 
This may be related to a more complex dependence of PD on the angle between the funnel axis and the line of sight.

We note that the orbital-average ultrasoft-state PD shows pronounced variations from one orbital cycle to another (Fig.~\ref{fig:time_variab}).
The changes could indicate secular variations of funnel parameters, but might as well be related to the switching between the dominating mechanisms responsible for the polarization production: reflection at the funnel walls and scattering within/on top of the funnel itself.
The latter possibility can be verified by comparing the spectra corresponding to different polarization levels: those coming from the scattering are expected to have no pronounced neutral iron line.
In our case, however, the normalisation of the line in time bin 5 and an upper limit in time bins 2+3 are comparable.

Time variations of PD may potentially be linked to the changes of the optical depth of the scattering material: higher Thomson optical depth $\tau$ leads to higher scattered fluxes.
Higher $\tau$ also leads to the reduction of PD, as per the increased role of (unpolarized) contribution of higher-order scatterings. 
Changes of $\tau$ may be the reason for the visible anti-correlation between the orbital fluence and measured PD (see Fig.~\ref{fig:time_variab}).
The PD drop by a factor of 1.5 between the time bins 2+3 and 5 corresponds to the transition from an optically thin medium to the medium of Thomson optical depth $\tau\approx0.35$.
We can then obtain the average bolometric luminosity of the system using the ultrasoft state flux estimate from \citet{Hjalmarsdotter2009} (with updated distance estimate and when accounting for the scattered fraction corresponding to $\tau=0.35$).
We obtain the intrinsic luminosity estimate $L_{\rm int}\approx1.4\times10^{39}$~erg~s$^{-1}$.
This estimate is close to the Eddington limit for He-rich material if the compact object has mass $\lesssim5$ solar masses, in line with previous results \citep{Zdziarski2013,KoljonenMaccarone2017}.

At the same time, the spectra corresponding to the high flux phases (orbital phase 0.5) show strong suppression of the iron line, as compared to the low fluxes (phase 0, see Appendix~\ref{app:ixpe}).
This may indicate systematic changes between scattering (phase 0.5) and reflection (phase 0) mechanisms at different orbital phases, attributed to the changing viewing angle of precessing funnel.
The observed high-flux PD=$11.2\pm0.4$\% corresponds to the scattering angle (inclination) in this phase $i=26\fdg7\pm0\fdg5$.

The polarimetric properties discussed above imply that during the ultrasoft state, when the jet is quenched, the funnel interior and the channel above it become filled with matter, coming from the WR wind and perhaps from the inner walls of the outflow \citep[this picture is similar to the jet cocoon scenario,][]{Koljonen2018}.
The amount of material along the funnel axis in this state is much larger than in the hard state, when the powerful jet cleared the material within the funnel beam (and further away above it) and the observer saw only the light reflected from the inner walls.
The drop of polarization during the intermediate (minor flaring) state might then be explained by the accumulation of matter close at the inner boundaries of the funnel and replacement of the reflection at the sharp inner boundary by the reflection within the volume close to this boundary (see Fig.~\ref{fig:pol_funnel}).
The effects of matter ionisation may also play a role in the PD reduction: in this case, the funnel becomes partially transparent to the reflected radiation (but not the incident one).

\section{Summary}

We have studied the X-ray polarimetric properties of the X-ray binary Cyg X-3 during the ultrasoft state.
The high PD=$11.9\pm0.5$\% and PA=$94\degr\pm1\degr$ indicate that the central source is covered by a thick envelope, similar to the previously observed hard and intermediate states.
The polarization of the continuum radiation can arise from a single scattering off the optically thin medium located along the funnel axis.
This is different from the scenario for the other states, where the X-ray continuum is dominated by the reflection from funnel walls.

We show that the PD in this case can serve as a direct probe of the orbital inclination, $i=27\fdg5\pm0\fdg5$.
We discuss the changes of the orbit-average and phase-dependent PDs and find that this scenario can be realised in the phases with highest flux, with PD=$11.2\pm0.4$\%, leading to the conservative inclination estimate $26\degr<i<28\degr$.

Changes of the orbital-average PD may be related to the optical depth variations of scattering matter.
The transition between the hard and ultrasoft states may in turn be caused by the filling of the funnel by the matter (coming either from the funnel walls or from the enhanced WR wind), linked to the jet quenching.
The major flaring state can then be related to the matter ejection from the funnel, its eventual emptying and transition of the source to the hard state.
This closes the loop of the correlated radio and X-ray activity in Cyg X-3.

\begin{acknowledgements}
The Imaging X-ray Polarimetry Explorer (IXPE) is a joint US and Italian mission. The US contribution is supported by the National Aeronautics and Space Administration (NASA) and led and managed by its Marshall Space Flight Center (MSFC), with industry partner Ball Aerospace (contract NNM15AA18C).  The Italian contribution is supported by the Italian Space Agency (Agenzia Spaziale Italiana, ASI) through contract ASI-OHBI-2022-13-I.0, agreements ASI-INAF-2022-19-HH.0 and ASI-INFN-2017.13-H0, and its Space Science Data Center (SSDC) with agreements ASI-INAF-2022-14-HH.0 and ASI-INFN 2021-43-HH.0, and by the Istituto Nazionale di Astrofisica (INAF) and the Istituto Nazionale di Fisica Nucleare (INFN) in Italy.  This research used data products provided by the IXPE Team (MSFC, SSDC, INAF, and INFN) and distributed with additional software tools by the High-Energy Astrophysics Science Archive Research Center (HEASARC), at NASA Goddard Space Flight Center (GSFC).

This research has been supported by the Academy of Finland grant 355672 (AV, VA) and the Vilho, Yrjö, and Kalle Väisälä foundation (SVF). 
ADM, FLM, EC, FM and PS are partially supported by MAECI with grant CN24GR08 “GRBAXP: Guangxi-Rome Bilateral Agreement for X-ray Polarimetry in Astrophysics”.
AAZ acknowledges support from the Polish National Science Center grants 2019/35/B/ST9/03944, 2023/48/Q/ST9/00138, and from the Copernicus Academy grant CBMK/01/24.
JPod and MD acknowledge the support from the Czech Science Foundation project GACR 21–06825X and the institutional support from the Astronomical Institute RVO:6798581.
DG and LR thank the staff of Lord's Bridge, Cambridge, for the their support in making the AMI observations.
\end{acknowledgements}


\bibliographystyle{yahapj}

\begin{appendix}

\section{\ixpe data}\label{app:ixpe}

\ixpe is the imaging polarimetry NASA/ASI mission launched in December 2021 \citep{Weisskopf2022}. 
It contains three grazing-incidence telescopes, each consisting of a mirror module assembly, which focuses X-rays onto a focal-plane polarization-sensitive gas pixel detector unit \citep[DU;][]{Soffitta21,Baldini21}.
\ixpe observed Cyg X-3 on 2024 June 2--3 (ObsID 03250301) with the net exposure time of 60~ks.

The data were processed with the {\sc ixpeobssim} package \citep{Baldini2022} version 31.0.1 using the CalDB released on 2024 February 28.
Source photons were extracted from a circular region centred on the source, with the radius of 80\arcsec.
Due to the brightness of the source, background subtraction was not applied and the unweighted approach was used \citep{Di_Marco_2022,Di_Marco_2023}.

We first perform a model-independent analysis of the orbital-averaged \ixpe polarimetric data using the \texttt{pcube} algorithm included in the \textsc{ixpeobssim} package, based on the formalism by \citet{Kislat2015}.
We compute the PD=$\sqrt{q^2+u^2}$ and the PA=$\frac{1}{2}\arctan (u/q)$, using the normalised Stokes parameters $q=Q/I$ and $u=U/I$.

To investigate the orbital dependence of the polarimetric data, we folded the observation with the quadratic ephemeris (model 2 from Table 2) of \citet{Antokhin2019}  using the \texttt{xpphase} method from the \textsc{ixpeobssim} package. 
The data were grouped in 9 bins and then the \texttt{pcube} algorithm was applied for each phase bin in the energy range 2--8~keV. 

Next, we utilised spectral information. 
The energy spectra were fitted simultaneously using the \textsc{xspec} package version 12.14.0 \citep{Arn96} using $\chi^2$ statistics and the version 20240101\_v013 of the instrument response functions.
The reported uncertainties are at the 68.3\% confidence level (1$\sigma$) unless stated otherwise. The spectropolarimetric analysis was performed using the weighted approach \citep{Di_Marco_2022}.

\begin{figure*}
\centering
\includegraphics[width=0.69\linewidth]{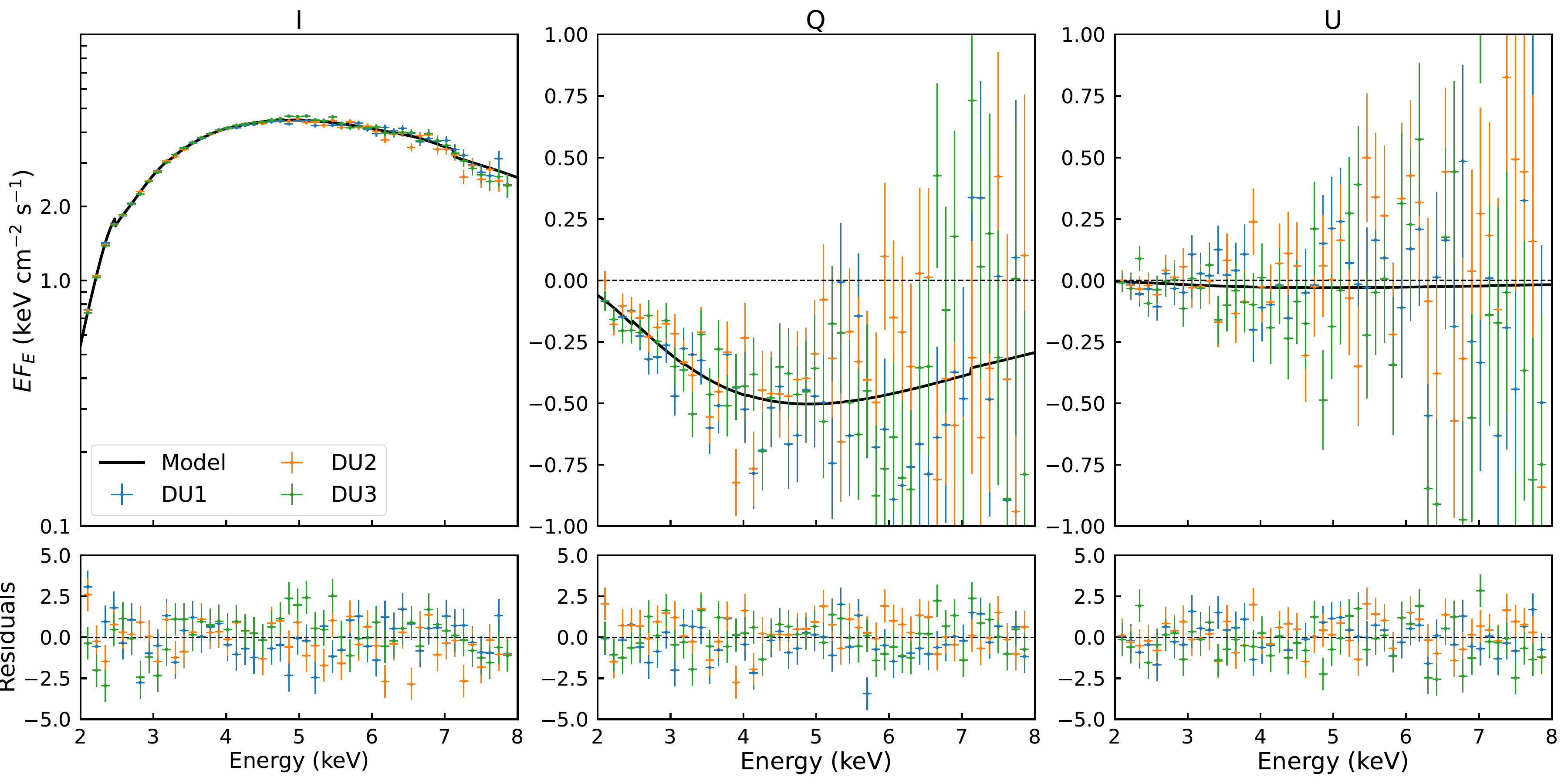}
\caption{Spectropolarimetric properties of Cyg X-3 corresponding to the high fluxes with the \ixpe count rate above 32~cnt~s$^{-1}$ (see Fig.~\ref{fig:time_variab}) around orbital phase 0.5.}
\label{fig:specpol_high} 
\end{figure*}

\begin{figure*}
\centering
\includegraphics[width=0.69\linewidth]{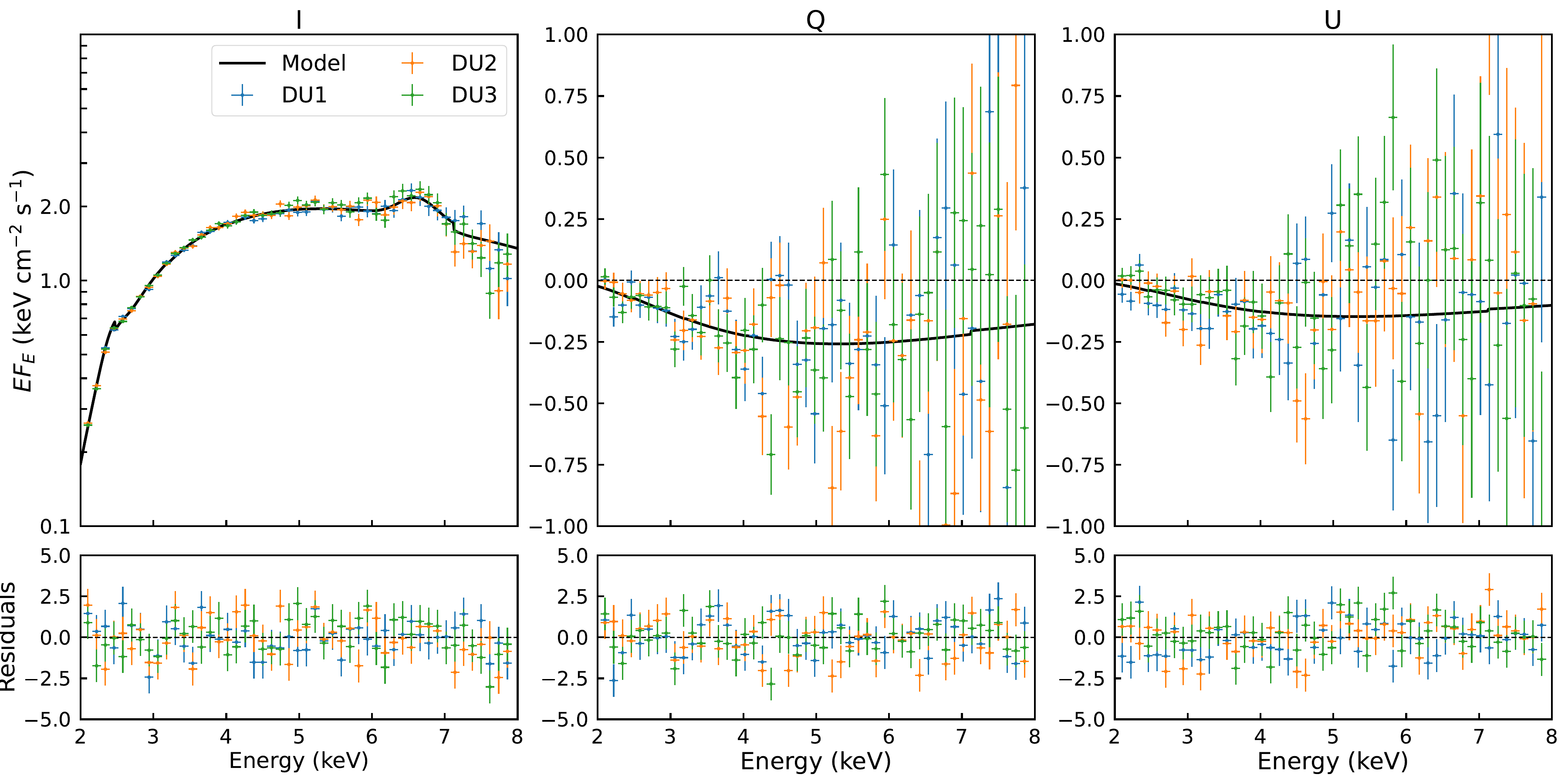}
\caption{Same as Fig.~\ref{fig:specpol_high} but for the low fluxes with the \ixpe count rate below 32~cnt~s$^{-1}$ (around orbital phase 0.0).}
\label{fig:specpol_low} 
\end{figure*}

We performed spectropolarimetric analysis of \mbox{Cyg X-3} at high and low flux levels corresponding to the time intervals when the \ixpe count rate was higher or lower than 32~cnt~s$^{-1}$ (see the light curve in top panel of Fig.~\ref{fig:time_variab}).
These levels roughly give the fluxes around orbital phase 0.5 and 0, respectively. 
The spectral model we used as the basis of the spectropolarimetric analysis, given the soft-state of the source, was \texttt{tbabs*(polconst*diskbb+gauss+gauss)}, where the two unpolarised gaussian lines are fixed at energies 2.4 and 6.6~keV with the width $\sigma$ of 0.15 and 0.25~keV, respectively.
The gaussians are responsible for the contribution of the unpolarized lines of \ion{Si}, \ion{S}, \ion{Fe}\,and other species identified by high-resolution spectroscopy \citep{Kallman2019}.
We also used a constant factor \texttt{const} to account for uncertainties in the effective area of different DUs.  
The best-fit results for the high and low fluxes are reported in Table~\ref{tab:spec_pol_flux}.
The deconvolved spectrum of Stokes parameters $I$, $Q$, and $U$ in $EF_E$ representation are shown in Figs.~\ref{fig:specpol_high} and \ref{fig:specpol_low}.

\begin{figure*}
\centering
\includegraphics[width=0.69\linewidth]{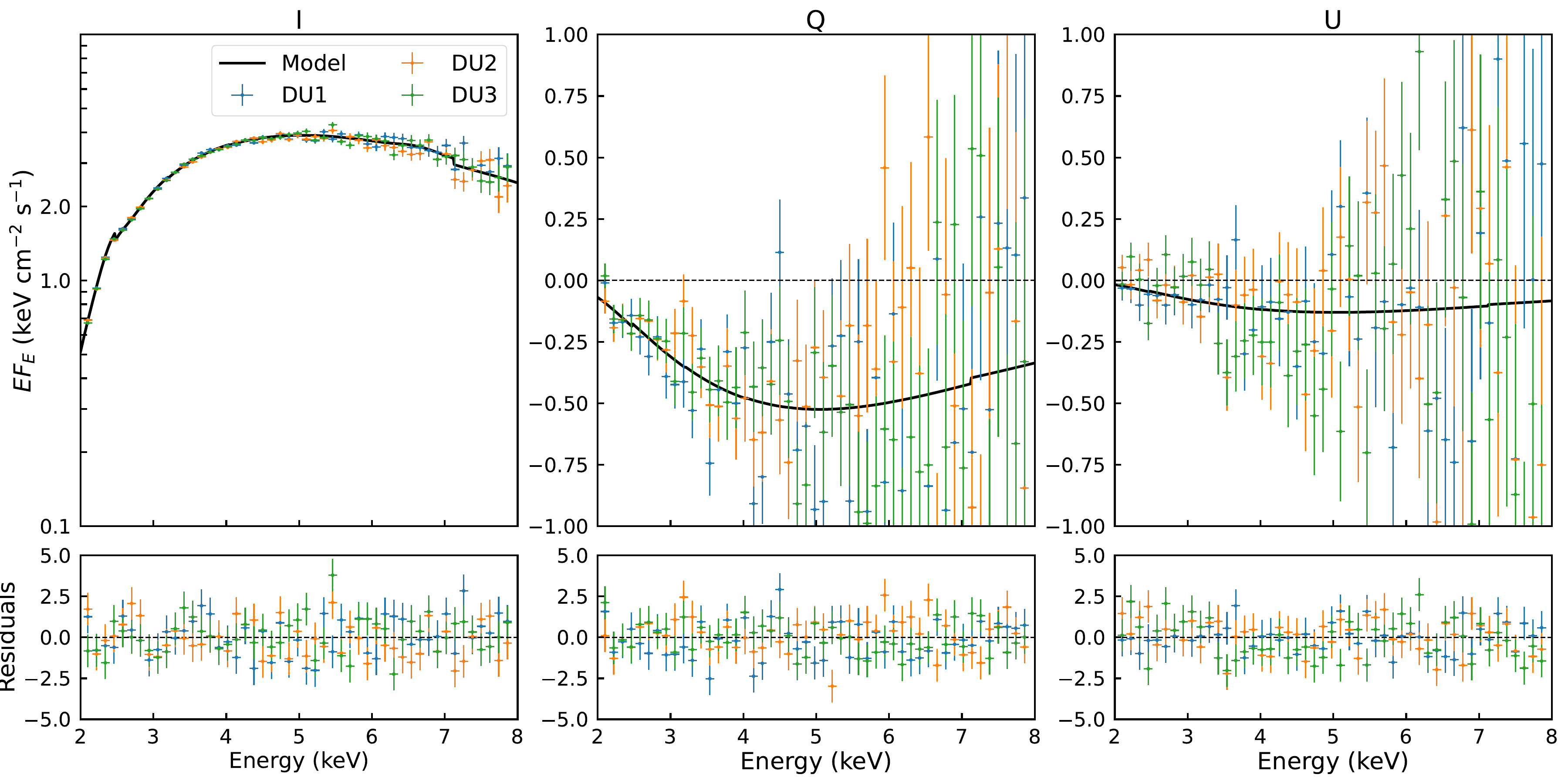}
\caption{Same as Fig.~\ref{fig:specpol_high} but averaged over the orbital phases, during the time bins 2 and 3 (see Fig.~\ref{fig:time_variab}).}
\label{fig:specpol_tbin23} 
\end{figure*}

\begin{figure*}
\centering
\includegraphics[width=0.69\linewidth]{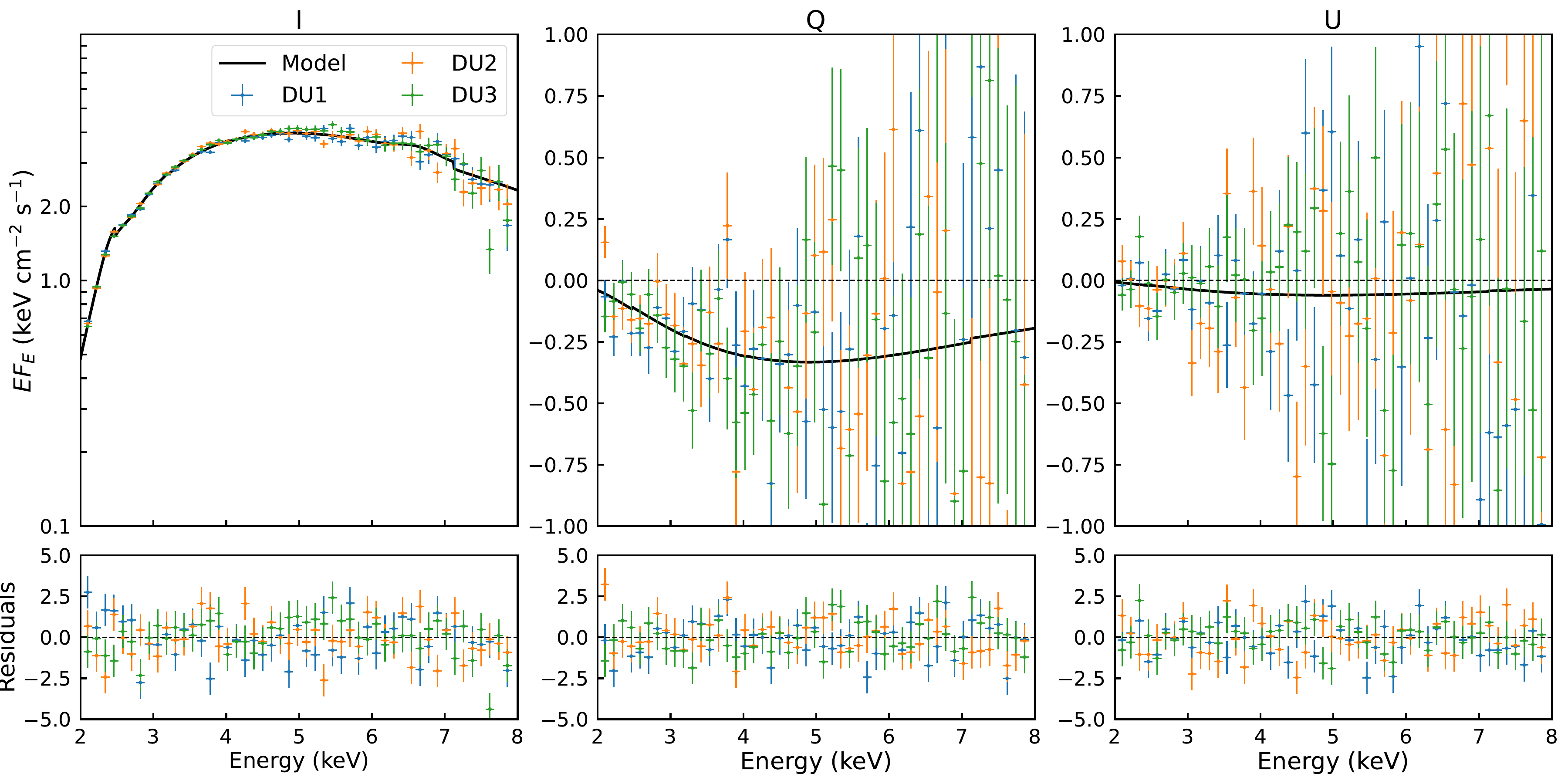}
\caption{Same as Fig.~\ref{fig:specpol_high} but averaged over the orbital phases, during the time bin 5 (see Fig.~\ref{fig:time_variab}).}
\label{fig:specpol_tbin5} 
\end{figure*}

Then, we performed the same spectropolarimetric analysis checking variation of the orbital-average polarization with time. 
In particular,  we performed the analysis putting together the data in the time bins 2 and 3 and in the time bin 5 (see Fig.~\ref{fig:time_variab}), corresponding to the highest and lowest PD. 
The best-fit results are given in Table~\ref{tab:spec_pol_bins}.
The deconvolved spectrum of Stokes parameters $I$, $Q$, and $U$ in $EF_E$ representation are shown in Figs.~\ref{fig:specpol_tbin23} and \ref{fig:specpol_tbin5}.

\begin{table}
\centering
\caption{Model parameters for the spectropolarimetric fit for the high and low fluxes.}
\begin{tabular}{cccc}
     \hline\hline
     Component  & Parameter  & Low flux & High flux \\
     \hline
     \texttt{tbabs}  & $N_{\rm H}$ ($10^{22}$~cm$^{-2}$) & $7.5\pm0.2$ & $7.0\pm0.1$ \\
     \texttt{diskbb} & $kT_{\rm in}$ (keV) & $1.8\pm0.4$ & $1.64\pm0.01$\\
     & norm & $32\pm$3 & $117\pm4$ \\
     \texttt{polconst} & PD (\%) & $15.2\pm0.9$ & $11.2\pm0.4$ \\
    & PA (deg) & $104.8\pm1.6$ & $91.7\pm0.9$ \\
    \texttt{gauss$_1$} & $E$ (keV) & [2.4] & [2.4] \\
    & $\sigma$ (keV) & [0.15] & [0.15]\\
    & norm ($\times10^{-2}$)& $2.6\pm0.6$ & $4.2\pm0.5$\\
    \texttt{gauss$_2$} & $E$ (keV) & [6.6] & [6.6] \\
    & $\sigma$ (keV) & [0.25] & [0.25] \\
    & norm ($\times10^{-3}$) & $6.0\pm1.6$ & $<2.5$\\
\texttt{const}     &  DU1 & [1] & [1] \\
    &  DU2 & $1.020\pm0.006$ & $1.016\pm0.002$\\
    &  DU3 & $1.006\pm0.006$ & $0.998\pm0.002$\\
     \hline
    & $\chi^2$/dof & 475/432 & 521/432 \\
     \hline
     \end{tabular}
\label{tab:spec_pol_flux}
\end{table}

\begin{table} 
\centering
\caption{Model parameters for the spectropolarimetric fit for the time bins 2+3 and 5 of Fig.~\ref{fig:time_variab}.}
\begin{tabular}{cccc}
     \hline\hline
     Component  & Parameter & Time bins 2+3 & Time bin 5 \\
     \hline 
     \texttt{tbabs}  & $N_{\rm H}$ ($10^{22}$~cm$^{-2}$)  & $6.40\pm0.12$ & $7.02\pm0.16$\\
     \texttt{diskbb} & $kT_{\rm in}$ (keV) & $1.74\pm0.19$ & $1.64\pm0.02$\\
     & norm & $76\pm4$ & $103\pm6$ \\
     \texttt{polconst} & PD (\%) &  $13.9\pm0.5$ & $8.5\pm0.7$ \\
    & PA (deg) &$97.0\pm1.1$ & $95\pm2$\\
    \texttt{gauss$_1$} & $E$ (keV) & [2.4] & [2.4]\\
    & $\sigma$ (keV) & [0.15] & [0.15] \\
    & norm ($\times10^{-2}$) &$3.1\pm0.6$ & $4.9\pm0.9$\\
    \texttt{gauss$_2$} & $E$ (keV) & [6.6] & [6.6] \\
    & $\sigma$ (keV) & [0.25] & [0.25] \\
    & norm ($\times10^{-3}$) & $<3.7$ & $3\pm2$\\
  \texttt{const}  &  DU1 & [1] & [1] \\
    &  DU2 & $1.007\pm0.004$ & $1.024\pm0.005$\\
    &  DU3 &$0.991\pm0.003$ & $1.003\pm0.005$\\
     \hline
    & $\chi^2$/dof & 479/432 & 491/432\\
     \hline
     \end{tabular}
\label{tab:spec_pol_bins}
\end{table}

\section{Radio behaviour at 15~GHz}\label{app:radio}

\begin{figure*}
\centering
\includegraphics[width=0.7\linewidth]{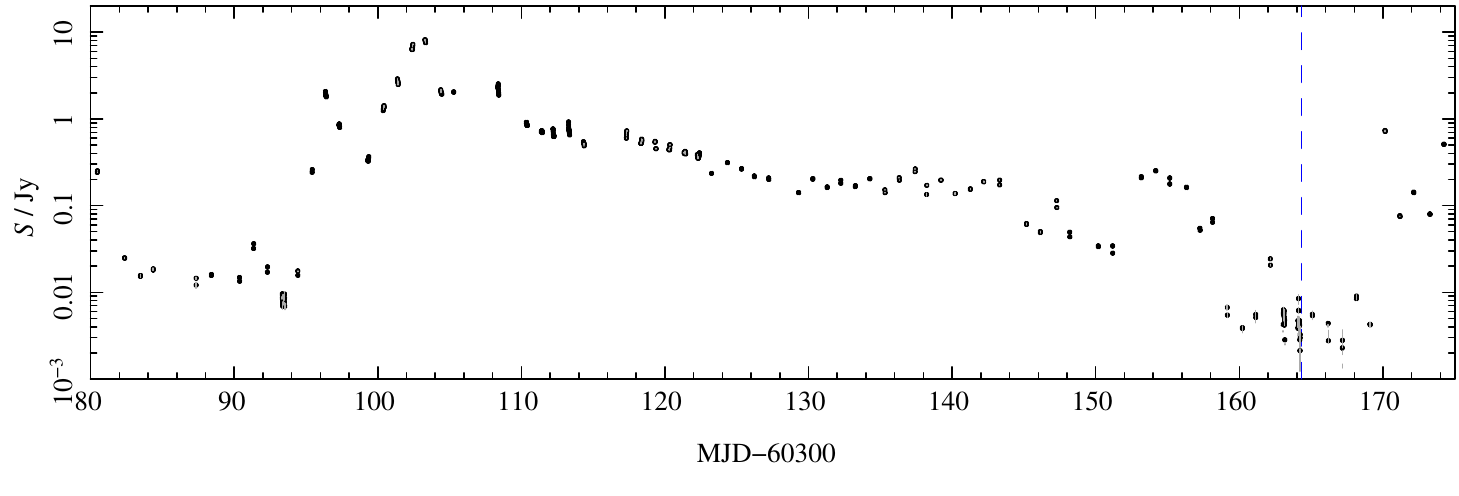}
\caption{AMI observations of Cyg X-3 at 15~GHz from  early March to mid June of 2024. Each point is the average flux density in a 10-min bin, with an error bar giving the statistical error. The dashed blue line indicates the \ixpe observation date.}
\label{fig:radio_ami1} 
\end{figure*}

\begin{figure*}
\centering
\includegraphics[width=0.7\linewidth]{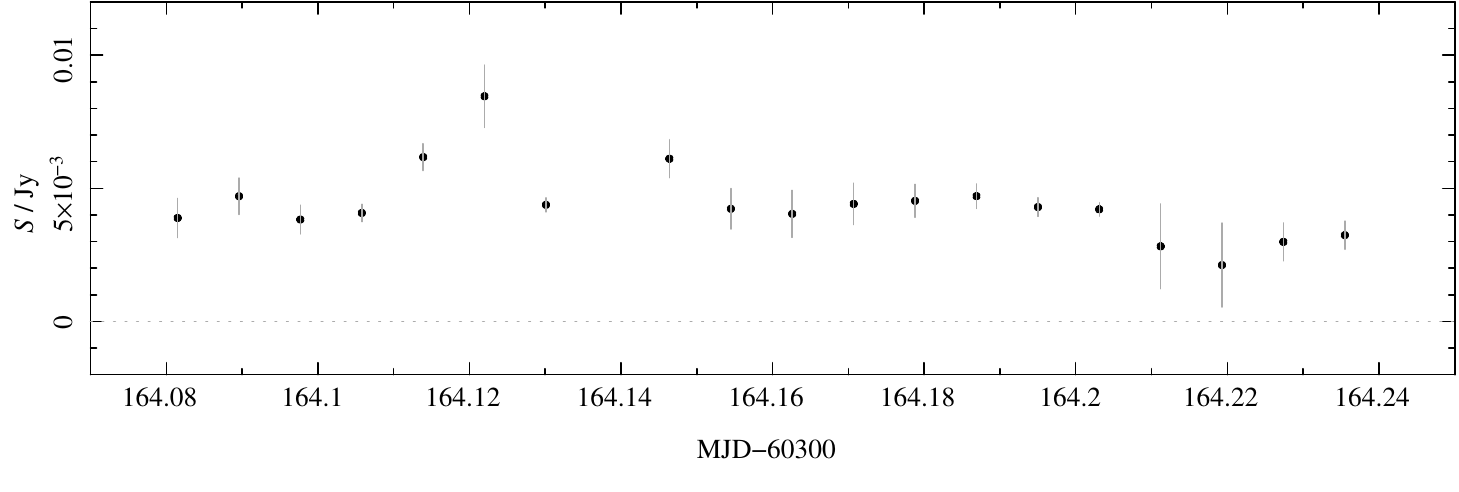}
\caption{Same as Fig.~\ref{fig:radio_ami1} but for the $\approx 4$~h (i.e. approximately one orbital period) long observation on 2024 June 3, simultaneous with \ixpe. }
\label{fig:radio_ami2} 
\end{figure*}

Cygnus X-3 is monitored at 15~GHz, usually daily, with the Large Array of Arcminute MicroKelvin Imager (AMI; \citealt{2008MNRAS.391.1545Z, 2018MNRAS.475.5677H}). AMI consists of eight 12.8-m antennas sited at the Mullard Radio Astronomy Observatory near Cambridge, UK. The AMI receivers cover the band from 13 to 18 GHz, and are of a single linear polarisation, Stokes $I+Q$. Observations are usually made daily, although some are missed due to adverse weather conditions (high winds or heavy rain), or technical issues. Also the number of antennas available varies due to technical issues. Analysis is done using custom software, \textsc{reduce\_dc} \citep{2013MNRAS.429.3330P}. 
Each observation consists of multiple 10-min scans of Cyg X-3, interleaved with short ($\sim 2$-min) observations of a nearby compact source, which were used for phase calibration. The flux density scale is set using nearby observations of 3C~286, which are usually made daily. The day-to-day flux density uncertainty is estimated at $\approx 5$\%. Usually short observations, with two 10~min scans are made daily, but sometimes longer observations are made. Figure~\ref{fig:radio_ami1} shows the AMI observations of \mbox{Cyg X-3} for several months before and weeks after the \ixpe observations. These show that at 15~GHz, \mbox{Cyg X-3} was in a faint state after a larger outburst in early April. These AMI observations include a longer, $\approx 4$ hour, observation of Cyg X-3 made on 2024 June 3, during the \ixpe observations, which is shown in Fig.~\ref{fig:radio_ami2}.
The source does not show any pronounced orbital variability in flux density.

\section{Orbital variability}\label{app:rvm}

To describe the orbital variability using the rotating vector model, we consider a funnel inclined at angle $\theta$ relative to the orbital axis and having azimuthal angle $\phi_0$ measured in the orbital plane from  the line connecting the black hole to the WR star.  
The orbital axis make angle $i$ to the line of sight. 
The PA of the scattered radiation can be written as $\chi=\chi_0+\chi_{\rm a}+\pi/2$, where $\chi_{\rm a}$ is the position angle on the sky of the orbital  angular momentum (measured from north to east) and $\chi_0$ is given by the rotating vector model (equation\,(30) in \citealt{Poutanen2020}):
\begin{equation} \label{eq:pa_rvm}
\tan \chi_0 = \frac{-\sin \theta\ \sin (\phi+\phi_0)}
{\sin i \cos \theta  - \cos i \sin \theta  \cos (\phi  + \phi_0) } .
\end{equation} 
The PD is determined by Eq.~\eqref{eq:pdmu} where the cosine of the scattering angle varies with the orbital phase as 
\begin{equation} 
\mu=\cos i \cos \theta + \sin i \sin \theta \cos(\phi+\phi_0).
\end{equation} 

The model thus has four parameters: $i$, $\theta$, $\chi_{\rm a}$, and $\phi_0$.
We fitted the ultrasoft state polarization data with this model obtaining $\chi^2=25.7$ for 8 dof. 
The best-fit parameters are $i=27\fdg7$,  $\theta=3\degr$, $\chi_{\rm a}=6\fdg5$, and $\phi_0=-120\degr$ (i.e. nearly in the direction of motion).  
This direction is not far from the position of the bow shock \citep{Antokhin2022} where an excess of the optical depth is expected.  
We did not try to get the confidence level on the parameters, because the fit is not acceptable.

\end{appendix}


\end{document}